\documentclass[a4paper,aps,prl,twocolumn]{revtex4}

\usepackage[dvips]{graphicx}
\usepackage{amsmath}
\usepackage{amssymb}
\usepackage{natbib}
\usepackage{psfrag}
\usepackage{wasysym}

\newcommand{\piper}{\Pi_{\perp}}
\newcommand{\pipar}{\Pi_{\parallel}}

\begin{document}

\title{The Granular Character of Particle Rafts}
\author{Pietro Cicuta$^1$ and Dominic Vella$^{2}$}
\affiliation{$^1$ Cavendish Laboratory and Nanoscience Centre, University of Cambridge, Cambridge CB3 0HE, U.~K.\\
$^2$ ITG, Department of Applied Mathematics and Theoretical Physics, University of Cambridge, Wilberforce Road, Cambridge, CB3 0WA, U.~K. }

\date{\today}

\begin{abstract}
We consider a monolayer of particles floating at a horizontal liquid-gas interface --- a particle raft. Upon compressing the monolayer in a Langmuir trough, the particles at first pack but ultimately the monolayer buckles out of the plane. We measure the stress profile within the raft at the onset of buckling and show for the first time that such systems exhibit a Janssen effect: the stress decays exponentially away from the compressing barriers over a length scale that depends on the width of the trough. We find quantitative agreement between the rate of decay and the simple theory presented by Janssen and others. This demonstrates that floating particle rafts have a granular, as well as elastic, character, which is neglected by current models. Finally, we suggest that our experimental setup may be suitable for exploring granular effects in two dimensions without the complications of gravity and basal friction.
\end{abstract}


\maketitle

A monolayer of densely packed particles floating at a liquid interface has many curious and potentially useful properties. At a coarse-grained scale, this particle-coated interface  is reminiscent of an elastic sheet since compression leads to a buckling instability \cite{vella04,luka} while in tension it may also fracture\cite{vella06}. Detailed experiments confirm a transition from liquid-like to solid-like behaviour as the particle concentration increases\cite{cicuta03}. At the scale of the individual particles, the complex shape of the liquid--gas interface has been shown to prolong the lifetime of particle-coated bubbles\cite{abkarian07}. Furthermore, colloidal particles are well known to be capable of stabilising emulsions and enable the formation of dry capsules or colloidosomes\cite{dinsmore02}. This rich variety of properties ensures that particle-coated interfaces have found application in diverse arenas from drug delivery\cite{tsapis02} to waste disposal by insects\cite{pike02}.

At the same time, an intensive and largely independent research effort has focused on understanding the stress state within granular materials. One of the classic papers in this field is by Janssen\cite{janssen95,sperl06} who considered the stress distribution in a granular silo. Janssen showed that the vertical pressure distribution in a silo filled with grains is not hydrostatic. Instead, the pressure saturates at a constant value which scales as the weight of grains over a vertical distance comparable to the width of the silo.

In this Letter we study the buckling instability of a particle raft, and address the question of what causes the raft to buckle. Previous experiments on very small (colloidal) particles and theoretical calculations have suggested that buckling occurs when the effective surface energy becomes zero\cite{aveyard00}. This is also the case for surfactant or lipid monolayers\cite{lipp98,takamoto01}.  In these experiments it is usual to give results in terms of the surface pressure $\Pi\equiv \gamma_c-\gamma$, in which $\gamma_c$ is the surface tension coefficient of the clean interface (i.e.~in the absence of particles) and $\gamma$ is the effective surface tension of the `contaminated' interface. Thus the orthodox view is that buckling occurs when $\Pi=\gamma_c$. Experimentally, it is conventional to measure the surface pressure using a Wilhelmy plate: a strip of material that is wetted by the underlying liquid and is inserted into the surface. By measuring the force exerted on the strip by the liquid interface, the surface pressure $\Pi$ may be inferred. Recently, it was suggested that this method of measurement may give distorted results when the interface is no longer fluid-like\cite{pocivavsek08}, e.g.~when a particle raft is very densely packed. Indeed it is well known that for viscoelastic or elastic layers, the surface pressure can be anisotropic, and is best regarded as a surface stress~\cite{cicuta04b}. Here, we present the first systematic  test of the orthodox view that $\Pi=\gamma_c$ at the onset of buckling and show that, as Pocivavsek \emph{et al.} intimated\cite{pocivavsek08}, the surface pressure is not uniform throughout the raft (albeit for a different physical reason). We find that the pressure measured at buckling by a Wilhelmy plate depends on the width of the trough and the separation of the barriers at this point. This observation cannot be explained using current elastic models and we show that it is a consequence of the granular stress state in the particle monolayer. Crucially, this demonstrates that such systems have granular characteristics that have not been considered previously.

\begin{figure}
\centering
\includegraphics[height=5cm]{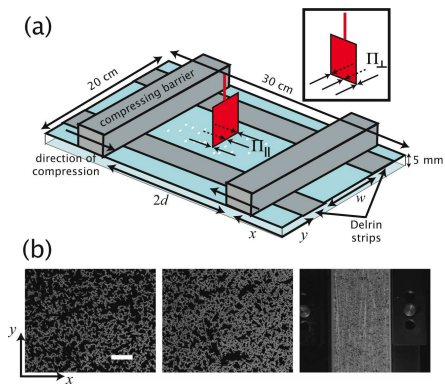}
\caption{The exerimental setup. (a) We use a Langmuir trough with barriers attached to a computer-controlled stepper motor. The width of the trough, $w$, is varied by moving the strips beneath the barriers. The computer is calibrated to infer the surface pressure $\Pi$ from the force on the Wilhelmy plate (red rectangle). In the configuration shown in the main figure it measures $\pipar$. To measure $\piper$, the plate is rotated by $90^\circ$ in the $(x,y)$-plane, as shown in the inset. Dimensions of the trough itself are for the NIMA 600. (b) Images showing the center of raft at three stages of compression. From left to right: initial state, onset of non-zero pressure and final, highly buckled, state. The scale bar represents $1\mathrm{~cm}$.}
\label{fig:setup}
\end{figure}

Our experimental setup is a monolayer of relatively large, athermal, particles\footnote{The particles used were made from Pliolite (Eliokem), a highly hydrophobic material, ground and sorted into different  samples\cite{vella04,vella06}. Here we used only particles with diameter in the range $112\mathrm{~\mu m}\leq d \leq 200\mathrm{~\mu m}$. These particles are big enough to be athermal but are small enough that surface tension dominates gravity.} at an air--water interface in a commercial Langmuir trough. Two different troughs were used: Minitrough (KSV, Finland) and NIMA 600 (NIMA, UK). Particles are added to the air--water interface in a petri dish of area  $\approx60\mathrm{~cm^2}$ until the addition of further particles would lead to interfacial buckling. This gives a reproducible `unit' of a large number of particles which are then transferred to the trough. The pliolite particles are so hydrophobic that none sink or are lost during this transfer. Before each run of the experiment, the interface is agitated manually to ensure that any particle clumps are broken up and the particles distributed evenly throughout the trough. The Langmuir troughs have two motorised barriers that are computer controlled so that the monolayer may be compressed symmetrically at constant barrier speed (the speed of compression is $100\mathrm{~\mu m/s}$\footnote{No dependence on compression speed was observed in the available range ($50-5000\mathrm{~\mu m/s})$.}).  A Wilhelmy plate placed at the center of the trough is used to measure (via supplied computer software) the surface pressure $\Pi$ as the monolayer is compressed. Two different orientations of this plate were used, as shown in fig.~\ref{fig:setup}a. In each experiment we use the sensor in one of these two configurations. We denote the surface pressures measured with the Wilhelmy plate by $\Pi_\perp$ and $\Pi_\parallel$ depending on whether the plate is perpendicular or parallel to the mobile barriers of the Langmuir trough. We are able to vary the width, $w$, of the Langmuir trough available to the monolayer by using specially cut strips of Delrin, placed in the subphase flush with the water surface, as shown in fig.~\ref{fig:setup}a.

This experimental system is different from that considered by Aveyard \emph{et al.}\cite{aveyard00} and others in three important ways. Firstly, in our system there is no long range repulsion between the particles (in fact, they are large enough that the attractive interaction energy from capillary forces is greater than thermal energy\cite{kralchevsky00}). The only repulsive interaction is the close-range steric repulsion. Secondly, the particles in this study are not perfect spheres and instead have rough shapes. Thirdly, our experimental system is unique because of its facility to vary the aspect ratio of the trough.

\begin{figure}
\includegraphics[height=7.5cm]{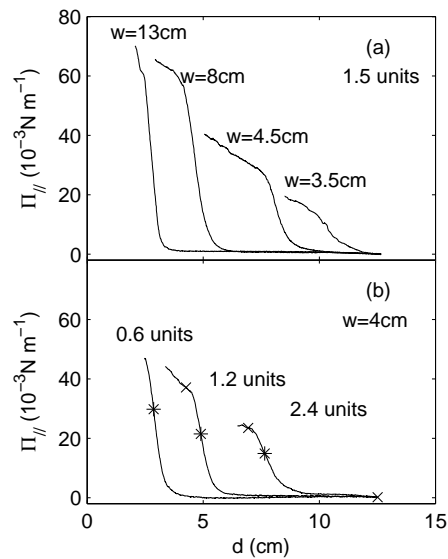}
\caption{Examples of experimentally measured surface pressure isotherms showing $\Pi$ as a function of the distance between the barriers and the sensor, $d$. $\Pi$ increases rapidly as $d$ decreases but then exhibits a kink (marked by $\times$ in b). This kink corresponds to the onset of buckling. Isotherms are shown for a range of trough widths, $w$, and for different numbers of particles spread on the surface. (a) With a fixed number of particles but decreasing trough width, the value of $\pipar$ measured at the onset of buckling decreases. (b) For fixed trough width but increasing number of particles,  the value of $\pipar$ at buckling decreases.}
\label{fig:isotherm}
\end{figure}

We measure the surface pressure as a function of the barrier separation, $2d$. This produces surface pressure isotherms such as those shown in fig.~\ref{fig:isotherm}. From these isotherms, we see that the surface pressure rises as the distance between the barriers decreases (i.e.~as the monolayer is compressed). After the onset of a non-zero pressure, the pressure  increases rapidly with further compression until at some critical barrier separation, $2d_c$, we observe a kink (marked by $\times$ in fig.~\ref{fig:isotherm}b). This is the compression at which the monolayer first buckles (monolayer collapse). It is unlikely that  a simple analytical form can be used to describe the measured isotherms. We focus instead on the measured value of the pressure at buckling, $\Pi^{buck}$. From the range of isotherms shown in fig.~\ref{fig:isotherm}, we see that $\Pi^{buck}$ depends on the width of the trough and also on the barrier separation at collapse, $2d_c$, which in turn depends on the number of particles spread on the interface\footnote{We measured the surface pressure in small regions of the interface where particles were excluded but which were accessible to any surfactant carried on the particles. We consistently found that $\Pi=0$ in such regions. Moreover, we note that when the number of particles is fixed but the aspect ratio of the trough at collapse is altered, the collapse pressure also changes. Both of these observations eliminate the possibility that the observed dependence on number of particles is due to surfactant contamination.}.

\begin{figure}
\centering
\includegraphics[width=8.5cm]{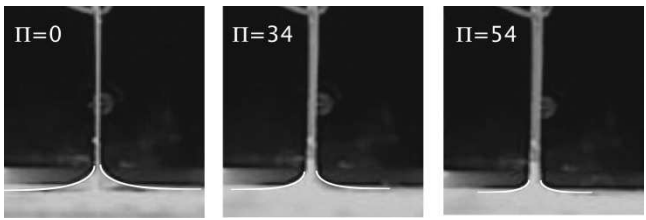}
\caption{Snapshots of the meniscus formed around the Wilhelmy plate compared to the theoretically predicted meniscus shape. Predicted shapes (solid curves) are obtained by solving the Laplace--Young equation \cite{landau} with the value of the surface tension corresponding to the measured surface pressure: $\gamma=\gamma_c-\Pi$.}
\label{fig:meniscus}
\end{figure}

Several authors have expressed concerns that the surface pressure measured by the Wilhelmy plate may not be an accurate representation of the effective surface tension of the interface in the vicinity of the plate\cite{pocivavsek08,kumaki88}. To test this directly, we took photographs of the meniscus shape around the Wilhelmy plate. For a given value of the surface pressure measured by the Wilhelmy plate, we determine the interface shape expected for a pure interface with this effective surface tension coefficient by solving the Laplace--Young equation with zero contact angle at the plate\cite{landau}. Such a comparison is shown in fig.~\ref{fig:meniscus} and shows excellent agreement between the theoretical and experimentally observed meniscus shapes. We therefore conclude that the measurements of the surface pressure by the Wilhelmy plate are consistent with the observed interface shape.

\begin{figure}
\includegraphics[height=7.5cm]{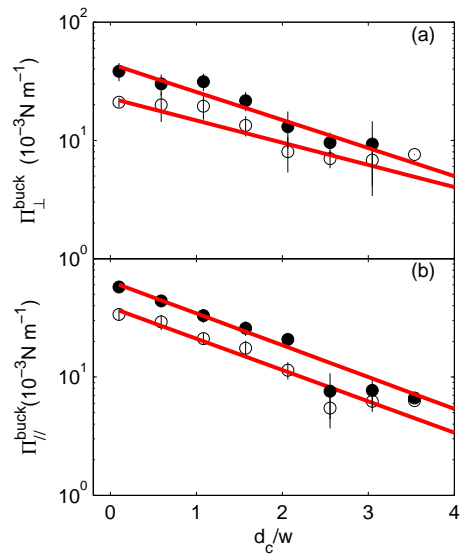}
\caption{Experimental data showing the pressure measured at the onset of buckling, $\Pi^{buck}$, as a function of the aspect ratio of the trough at this point, $d_c/w$. Results are shown for (a) $\piper$ and (b) $\pipar$. In each panel there are  two datasets, corresponding to two automated measures of $\Pi^{buck}$ from surface pressure isotherms: the inflection point (open symbols) and the point where the derivative falls to small values (filled symbols). Fits to the exponential decay of eq.~\eqref{pbuck} are shown with characteristic decay parameters $\lambda=1.8$ and 2.3 for $\piper$ and $\lambda=1.6$ for both datasets of $\pipar$. The prefactor in the exponential decay is sensitive to the definition of the buckling point and so is not considered here.  (A power law decay is not appropriate judging from the corresponding logarithmic plot and would not yield finite $\Pi^{buck}$ in the limit $d_c/w\rightarrow0$.)}
\label{fig:collapse}
\end{figure}

Having seen that the aspect ratio of the trough at collapse, $2d_c/w$,  influences the surface pressure at the onset of buckling $\Pi^{buck}$, we now present experimental results for a range of trough widths and number of particles at the surface. The pressure $\Pi^{buck}$ is defined as the kink in the surface pressure isotherm. To automate the detection of this point, we determine the  point of inflection of the isotherm, marked (*) in fig.~\ref{fig:isotherm}b, and the point where the derivative falls to small values, marked $(\times)$ in fig.~\ref{fig:isotherm}b. In fig.~\ref{fig:collapse} we show the results of these experiments presenting the pressure at buckling $\Pi^{buck}$ as a function of $d_c/w$. We performed 163 independent experiments, but for clarity of presentation we have averaged the results into bins that are equally spaced in $d_c/w$. These bins range from $d_c/w=0.1$ to $d_c/w=3.5$ --- a span of one and a half decades. Error bars represent the standard deviation of data values within each bin. We observe good collapse of data from a wide range of experiments in different troughs. This master curve suggests an exponential decay in $\Pi^{buck}$ with $d_c/w$. The prefactor of this exponential depends on the algorithm used to determine $\Pi^{buck}$ but the decay rate is insensitive to this. We therefore  focus on this decay rate as the most robust feature of the results presented in fig.~\ref{fig:collapse}.

We use ideas from the theory of granular elasticity \cite{jiang03} to explain the exponential decay presented in fig.~\ref{fig:collapse}.  Under the assumption that the material is elastic, and that there is no strain in the $y$-direction (perpendicular to the direction of compression, see fig.~\ref{fig:setup}a) we find that
\begin{equation}
\sigma_y=\nu \sigma_x
\label{jansenneq}
\end{equation}
where $\nu$ is the Poisson ratio\cite{ovarlez05}. We note  that \eqref{jansenneq} is simply the relationship posited by Jansenn between the vertical and horizontal pressure in a silo\cite{janssen95,sperl06}: here, the constant of proportionality $K=\nu$\cite{ovarlez05}. Furthermore, we note that this relationship should hold only close to buckling since prior to this particles can, and do, accommodate deformation by moving in the $y$ direction.

In our two-dimensional situation,  $\sigma_x=\pipar$ and $\sigma_y=\piper$. We now use a well-known balance of forces argument\cite{wambaugh08} to determine $\pipar(x)$. (Though $\pipar$ may in fact be a function of $y$, we consider only its $x$ dependence here.) To determine the behavior of $\pipar$ we consider an infinitesimal slice through the trough (taken at constant $x$) with width $w$. In equilibrium, the difference in pressure between the two faces of the slice must be balanced by frictional forces acting at the walls of the trough, denoted $\sigma_{xy}$, i.e.:
\begin{equation}
w\frac{d\pipar}{dx}=-2\sigma_{xy}.
\label{fbalance}
\end{equation}
 Assuming a Coulomb friction law and a friction coefficient $\mu$ we have that $\sigma_{xy}=\mu\piper=\mu\nu\pipar$. We may then integrate eq.~\eqref{fbalance}, which has solution:
\begin{equation}
\pipar(x)=\alpha\exp\bigl(-2\mu\nu x/w\bigr),
\end{equation} for some constant of integration $\alpha$.

We observe experimentally (see fig.~\ref{fig:setup}b) that buckling occurs first in the vicinity of, and parallel to, the compressing barriers. Under the assumption that this corresponds to a critical local pressure $\pipar=\gamma_c$ then the pressures measured by the Wilhelmy plate should be
\begin{equation}
{\pipar}^{buck}=\gamma_c\exp\bigl(-2\mu\nu d_c/w\bigr),\quad {\piper}^{buck}=\nu{\pipar}^{buck}.
\label{pbuck}
\end{equation}
In this relationship the constants $\mu$ and $\nu$ are the only unknowns. Two values of $\nu$ have been proposed for the case of perfect discs: $\nu=1/3$\cite{clegg08} and $\nu=1/\sqrt{3}$\cite{vella04}. We measured the friction coefficient $\mu$ directly; the angle of friction between Pliolite and the Delrin strips was measured to be $28^\circ$ so that $\mu\approx0.53$. Using these values we expect $\log(\pipar^{buck}/\gamma_c)=-(d_c/w)/\lambda$ with $\lambda=2.83$ or $\lambda=1.63$ using $\nu=1/3$ and $1/\sqrt{3}$, respectively. These predicted lines have a slope of the same order of magnitude as that measured experimentally. This suggests that the pressure decrease observed at the center of the raft relative to the value near the compressing barriers is indeed induced by granular effects. We note that taking $\nu=1/\sqrt{3}\approx0.577$ gives a significantly better account of the experimentally measured slope than the value $\nu=1/3$. Motivated by eq.~\eqref{jansenneq}, we calculate the ratio of $\piper^{buck}(d=0)$ and $\pipar^{buck}(d=0)$ to determine an independent estimate of the Poisson ratio, $\nu$. This ratio gives $\nu=0.63\pm0.05$. $\nu$ may also be estimated from the measured values of $\lambda$. This procedure gives $\nu=0.52\pm0.09$. These values are consistent with one another and with the value for rigid discs\cite{vella04}. 

Previously, particle rafts have been modelled as elastic sheets \cite{vella04,luka,cicuta03,pocivavsek08}. In this work we have shown that these approaches are insufficient --- even very recent elastic calculations\cite{pocivavsek08} do not predict the exponential decay reported here. An alternative interpretation of the last three data points in figure \ref{fig:collapse}b is that $\Pi^{buck}$ plateaus at a constant value for large values of $d_c/w$. However, even in this instance the granular character of the raft remains clear. The granular character of such rafts is therefore, in addition to their elastic character, a vital ingredient that must be considered when modelling them in the range of industrial\cite{dinsmore02} and biological\cite{pike02} settings in which they arise. 

Recently, rafts of bubbles\cite{lundberg08} and millimetric beads\cite{twardos06} have proven useful for understanding shear-banding and glassy dynamics. In the same way, the present system of rough, floating grains may prove useful as a model system for studying the influence of friction in two-dimensional granular materials. Furthermore, using our system to study the jamming transition would remove the possible effects of basal friction\cite{majmudar07} and allow confirmation of recent theoretical predictions for the influence of inter-particle attraction\cite{lois08} by exploiting the small capillary attraction\cite{kralchevsky00}. In addition, the Wilhelmy plate provides a convenient method of probing the stress state within the medium.

\textit{Acknowledgements}. We thank R.~Blumenfeld, E.~Cerda and L.~Mahadevan for useful discussions. D.V.~is supported by the 1851 Royal Commission.

\end{document}